\documentclass{article}
\usepackage{comment}
\usepackage{arxiv}
\usepackage[utf8]{inputenc} % allow utf-8 input
\usepackage[T1]{fontenc}    % use 8-bit T1 fonts
\usepackage{hyperref}       % hyperlinks
\usepackage{url}            % simple URL typesetting
\usepackage{booktabs}       % professional-quality tables
\usepackage{amsfonts}       % blackboard math symbols
\usepackage{nicefrac}       % compact symbols for 1/2, etc.
\usepackage{microtype}      % microtypography
\usepackage{lipsum}		% Can be removed after putting your text content
\usepackage{graphicx}
\usepackage{natbib}
\usepackage{doi}
\usepackage{booktabs}       % professional-quality tables
\usepackage{amsfonts}       % blackboard math symbols
\usepackage{nicefrac}       % compact symbols for 1/2, etc.
\usepackage{microtype}      % microtypography
\usepackage{fancyhdr}       % header
\graphicspath{{media/}}     % organize your images and other figures under media/ folder
\usepackage{array}
\usepackage{subfig}
\usepackage{xcolor}
\title{Artificial Intelligence for All? Brazilian Teachers on Ethics, Equity, and the Everyday Challenges of AI in Education}

\author{\href{https://orcid.org/0000-0000-0000-0000}{\includegraphics[scale=0.06]{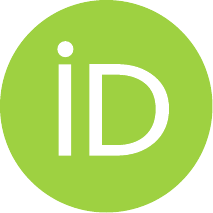}\hspace{1mm}Bruno Florentino}\thanks{These authors contributed equally and share first authorship.} \\
	Department of Computer Science\\
	University of São Paulo\\
    São Carlos, São Paulo, Brazil\\
	\texttt{brunorf1204@usp.br} \\
	\And
	\href{https://orcid.org/0000-0000-0000-0000}{\includegraphics[scale=0.06]{orcid.pdf}\hspace{1mm}Camila Sestito}\\
	Department of Computer Science\\
    Federal University of Technology-Paraná (UTFPR)\\
	Cornélio Procópio, Paraná, Brazil \\
	\texttt{camilasestito@utfpr.edu.br} \\
    \And
	\href{https://orcid.org/0000-0000-0000-0000}{\includegraphics[scale=0.06]{orcid.pdf}\hspace{1mm}André de Carvalho}\\
	Department of Computer Science\\
	University of São Paulo\\
    São Carlos, São Paulo, Brazil\\
	\texttt{andre@icmc.usp.br} \\
    \And
	\href{https://orcid.org/0000-0000-0000-0000}{\includegraphics[scale=0.06]{orcid.pdf}\hspace{1mm}Wellington Cruz}\\
	Educational President\\
	Instituto Significare\\
	São Paulo, São Paulo, Brazil\\
	\texttt{prof.well@significare.org.br} \\
    \And
	\href{https://orcid.org/0000-0000-0000-0000}{\includegraphics[scale=0.06]{orcid.pdf}\hspace{1mm}Robson Bonidia*}\thanks{Corresponding author: bonidia@utfpr.edu.br} \\
	Department of Computer Science\\
    Federal University of Technology-Paraná (UTFPR)\\
	Cornélio Procópio, Paraná, Brazil\\
	\texttt{bonidia@utfpr.edu.br} \\
}
\renewcommand{\shorttitle}{AI for All? Brazilian Teachers on Ethics, Equity, and the Everyday Challenges}
\hypersetup{
pdftitle={AI for All?},
pdfsubject={q-bio.NC, q-bio.QM},
pdfauthor={Anonymized information for review},
pdfkeywords={First keyword, Second keyword, More},
}
\begin{document}
\maketitle

%%%%%%%%%%%%%%%%%%%%%%%%%%%%%%%%%
%%%%%%%%%%%%%%%%%%%%%%%%%%%%%%%%%

\begin{abstract}
	High-quality early education benefits several other areas, including the environment, health, and social stability. Unfortunately, not everyone has access to high-quality education, which can widen the divide between countries and within countries. Artificial Intelligence (AI) can be a powerful tool for both educational improvement and equality. This study examines the perceptions of Brazilian K-12 education teachers regarding the use of AI in education, specifically General Purpose AI, which encompasses Generative AI, and their views on its benefits, challenges, and ethical implications. This investigation employs a quantitative analysis approach, extracting information from a questionnaire completed by 346 educators from various regions of Brazil regarding their AI literacy and use. Educators vary in their educational level, years of experience, and type of educational institution. The analysis of the questionnaires shows that although most educators had only basic or limited knowledge of AI (80.3\%), they showed a strong interest in its application, particularly for the creation of interactive content (80.6\%), lesson planning (80.2\%), and personalized assessment (68.6\%). The potential of AI to promote inclusion and personalized learning is also widely recognized (65.5\%). The participants emphasized the importance of discussing ethics and digital citizenship, reflecting on technological dependence, biases, transparency, and responsible use of AI, aligning with critical education and the development of conscious students. Despite enthusiasm for the pedagogical potential of AI, significant structural challenges were identified, including a lack of training (43.4\%), technical support (41.9\%), and limitations of infrastructure, such as low access to computers, reliable Internet connections, and multimedia resources in schools. The study shows that Brazil is still in a \textit{bottom-up} model for AI integration, missing official curricula to guide its implementation and structured training for teachers and students. This study also shows the strong benefits of using AI to support teaching, reduce workload, and personalize learning. However, its effective implementation depends on integrated public policies, adequate teacher training, and equitable access to technology, promoting ethical, inclusive, and contextually grounded adoption of AI in Brazilian K-12 education. The recommendations of this study include the implementation of free, online, continuous professional development programs aligned with teacher realities, the incorporation of ethics and digital citizenship into curricula, and the investment in technology infrastructure and technical support.
\end{abstract}

% keywords can be removed

%%%%%%%%%%%%%%%%%%%%%%%%%%%%%%%%%
%%%%%%%%%%%%%%%%%%%%%%%%%%%%%%%%%

\section{Introduction}

New technologies can transform lifestyles and reconfigure relationships between different sectors of society. However, their development and application prioritize countries in the Global North \citep{norte_sul2}. This has recently occurred with Artificial Intelligence (AI), in particular with models and systems based on Generative Artificial Intelligence (GAI), which, being primarily designed to meet the demands and contexts of countries of the global North, often overlook the specific needs and challenges faced by countries of the Global South \citep{global_north}.

This study analyzes the educational situation in Brazil, a country that faces significant challenges in public access to technology, ranking 44th among 133 economies in a global index on this issue \citep{portulans2024nri}. This is mainly due to social and structural inequalities in the country, for example, \cite{gozzi2024bridging} identified a correlation between internet speed and wealth indicators, revealing that populations with higher purchasing power tend to experience significantly better connectivity. The greatest disparity was observed in the city of Rio de Janeiro, where the speed of the Internet varies considerably between the favela regions and the rest of the city.

The aforementioned problems did not emerge with the advent of AI or GAI, but tend to become more pronounced when discussing their integration into society \citep{norte_sul2}. AI refers to a set of algorithms capable of learning from data and, based on that, performing tasks related to such data. Specifically, GAI is a type of AI trained in large volumes of data, such as texts, images, audio, and videos, which it uses to generate new outputs in the same formats, including new texts, images, and other multimodal content \citep{zhang_ai_benefits}.

In the educational context, several studies have highlighted both potential benefits and risks associated with the use of GAI, which must be carefully considered and balanced during its implementation \cite{opport_canada}. Among the positive impacts of GAI in education, the benefits for educators and students stand out. From the teachers' perspective, GAI can serve as a supportive tool, helping in the creation of teaching materials and accelerating the instructional design process \citep{art_ger, opport_canada, transforming_ai_benefits, zhang_ai_benefits, qiu_ai_benefits, berg_ai_benefits}. In addition, it can contribute to the performance of administrative tasks, such as drafting project proposals and recommendation letters \citep{opport_canada}.

Another relevant benefit lies in the automated correction of assignments, allowing instant feedback for students \citep{art_ger, opport_canada, zhang_ai_benefits, qiu_ai_benefits}. This functionality can be used by teachers to optimize the evaluation time and by students to test and consolidate their knowledge. In addition, GAI can generate visual content, such as images, videos, and simulations, facilitating the understanding of complex concepts \citep{art_ger, zhang_ai_benefits}. It is also possible to create personalized texts and activities based on previous student performance, promoting more personalized learning \citep{art_ger, transforming_ai_benefits, zhang_ai_benefits}.

However, many educators express concerns regarding the use of GAI in educational settings. One of the main concerns is the consistency and quality of the responses generated \citep{art_ger, transforming_ai_benefits, qiu_ai_benefits, berg_ai_benefits}. Studies, for example, show that the GPT-4 model achieves only 77\% predictive accuracy when returning the correct code in programming tasks \citep{art_ger}. In addition, there is evidence that the performance of the model in advanced mathematics remains unsatisfactory \citep{transforming_ai_benefits}.

Another recurring concern involves the biases embedded in the models, as the generated content tends to reflect the patterns of the data used during training \citep{art_ger, opport_canada, transforming_ai_benefits, berg_ai_benefits}. This problem is further exacerbated by the low interpretability of these technologies, which makes it difficult to understand the criteria used during the response generation process \citep{art_ger, transforming_ai_benefits}.

One more discussed point is that, since GAI's results are based on its training data, the content produced tends to represent more of a "pseudo-imagination" than something genuinely new and authentic \citep{art_ger}. In other words, it is a synthesis of what has been learned, rather than a true innovation. Therefore, its application in education should be carried out with caution, to avoid generic and repetitive outcomes that can discourage both teachers' creativity in teaching and the authenticity of students in presentations and original productions \citep{art_ger}. In this context, educators also report difficulties in distinguishing content genuinely produced by students from that generated by GAI, raising important debates about how to ensure academic integrity in assessment methods \citep{art_ger, opport_canada}.

Beyond academic issues, the emotional impact of introducing these tools is also being discussed \citep{qiu_ai_benefits}. Although GAI can provide immediate responses, it lacks the capacity for emotional communication and personalization that only a human teacher can offer. This human mediation is essential to provide emotional support to students \citep{qiu_ai_benefits}. In a scenario where students start to prefer AI interactions rather than turning to teachers, there is a risk of weakening the student–teacher relationship, making it more difficult for educators to identify students’ emotional and pedagogical needs \citep{qiu_ai_benefits}.

Given this scenario, which encompasses both the inequalities in access to technology in Brazil and the potential impacts of GAI in K-12 education, it becomes essential to understand the perceptions of Brazilian educators of this technology, its applicability, and its potential benefits within a national context. This type of research also provides information on Latin America, as Brazil is one of the most developed countries in the region \cite{HKTDC2025Brazil}.

To this end, we propose a quantitative approach aimed at investigating the opinions of teachers of K-12 education in Brazil about AI and GAI, exploring their expectations, perceived impacts, and the challenges associated with its implementation.  To our knowledge, this is the first nationwide survey designed to identify the actual needs of classroom teachers about the use of AI\footnote{Anonymized information for review.}. Therefore, the following research questions (RQ) were formulated:

\begin{itemize}
    \item \textbf{RQ1:} Considering the Brazilian context, do educators believe AI can bring benefits to education?
    \item \textbf{RQ2:} If so, which benefits are considered most relevant by educators?
    \item \textbf{RQ3:} Regarding student education, do educators consider it necessary to use the school environment to address ethical issues related to AI?
    \item \textbf{RQ4:} Concerning teacher training, are there sufficient materials and resources to promote the continuous professional development of teachers regarding AI use?
    \item \textbf{RQ5:} What are the main challenges in implementing AI in Brazilian K-12 education?    
\end{itemize}

Understanding these aspects is crucial for any proposal to adopt these technologies in basic education in Brazil and Latin America, respecting educators' perceptions, identifying their support needs, and considering the challenges reported in relation to implementation.

%%%%%%%%%%%%%%%%%%%%%%%%%%%%%%%%%
%%%%%%%%%%%%%%%%%%%%%%%%%%%%%%%%%

\section{Methodology}

This section discusses the methodology applied in this study, covering the foundations on which the questionnaire was developed, the experimental setup, and a statistical description of the sample obtained.

%%%%%%%%%%%%%%%%%%%%%%%%%%%%%%%%%
%%%%%%%%%%%%%%%%%%%%%%%%%%%%%%%%%

\subsection{Framework of Teachers' Digital Competencies}

Continuing professional development for teachers is a central topic in discussions about education, especially in the context of constant technological transformation. The ability to articulate and integrate new technologies into the school routine enables a contemporary and often more efficient approach to the teaching–learning process. In this regard, the Brazilian Ministry of Education published the \textbf{Framework of Teachers' Digital Competencies} \citep{inep2025censo_escolar}, aiming to stimulate self-development, promote reflection, and support the ongoing professional development of education professionals, taking into account the use of digital technologies in K-12 education.

This framework is structured into three dimensions that encompass a total of ten competencies. The first dimension is \textbf{"Teaching and Learning with the Use of Digital Technologies"}, focused on understanding and applying principles that facilitate the integration of digital technologies into pedagogical strategies. This dimension also includes improvements in content creation, data management, and the promotion of inclusive classroom practices. 

In this context, AI-based technologies are not only welcomed but also highly relevant, as they align with established objectives and contribute to significant improvements in teaching practice. This dimension comprises four specific competencies: pedagogical practice, curation and creation, data analysis, and inclusive practice.

The second dimension is \textbf{``Digital Citizenship''}, considered essential to prepare students for ethical, critical, and responsible engagement in the digital environment. This dimension involves understanding and applying principles related to safe and ethical behavior in the use of digital technologies. It encompasses everything from the protection of personal data and privacy to the critical use of information, including awareness of the impacts of excessive use of technology on student mental health and well-being. This dimension comprises three competencies: responsible use, safe use, and critical use.

Finally, the third dimension refers to \textbf{``Professional Development''}, which emphasizes the importance of continuously empowering teachers in the face of constant technological and pedagogical innovations. This dimension values the use of digital resources that support ongoing professional development, collaborative work, and the efficient management of pedagogical practices. Continuous professional development is essential for educators to incorporate new tools and approaches, stay up-to-date, and prepare for the challenges of teaching in the digital era. This dimension comprises three competencies: continuous professional development; communication and collaboration; and the use of digital resources for management.

%%%%%%%%%%%%%%%%%%%%%%%%%%%%%%%%%
%%%%%%%%%%%%%%%%%%%%%%%%%%%%%%%%%

\subsection{Experiment Design}

To address the research questions, a quantitative approach was employed to assess teachers' perceptions on the topic. A questionnaire consisting of 35 items was constructed to explore the thematic dimensions of the previously mentioned Framework of Teachers' Digital Competencies. These elements were derived from an initial questionnaire based on a literature review, which was initially administered to four educators from different regions of the country. A focus group between researchers and educators led to the final version of the material, which can be accessed in the Final Form – Artificial Intelligence in the Classroom\footnote{Anonymized information for review.}%\url{https://forms.gle/aTRJ4LE257EugDR56}}.

Data were collected from October 28, 2024, to February 28, 2025, using an online form (Google Forms), with the aim of a wide reach in the different regions of Brazil. The sample consisted of 346 teachers from various regions. Considering the total population of approximately 2.4 million active teachers in Brazilian basic education, according to the 2024 School Census \citep{inep2025censo_escolar}, this provides a confidence level of 90\% with a margin of error of 5\%.

%%%%%%%%%%%%%%%%%%%%%%%%%%%%%%%%%
%%%%%%%%%%%%%%%%%%%%%%%%%%%%%%%%%

\subsection{Sample}

\begin{figure}[!htp]
    \centering
    \includegraphics[width=0.95\linewidth]{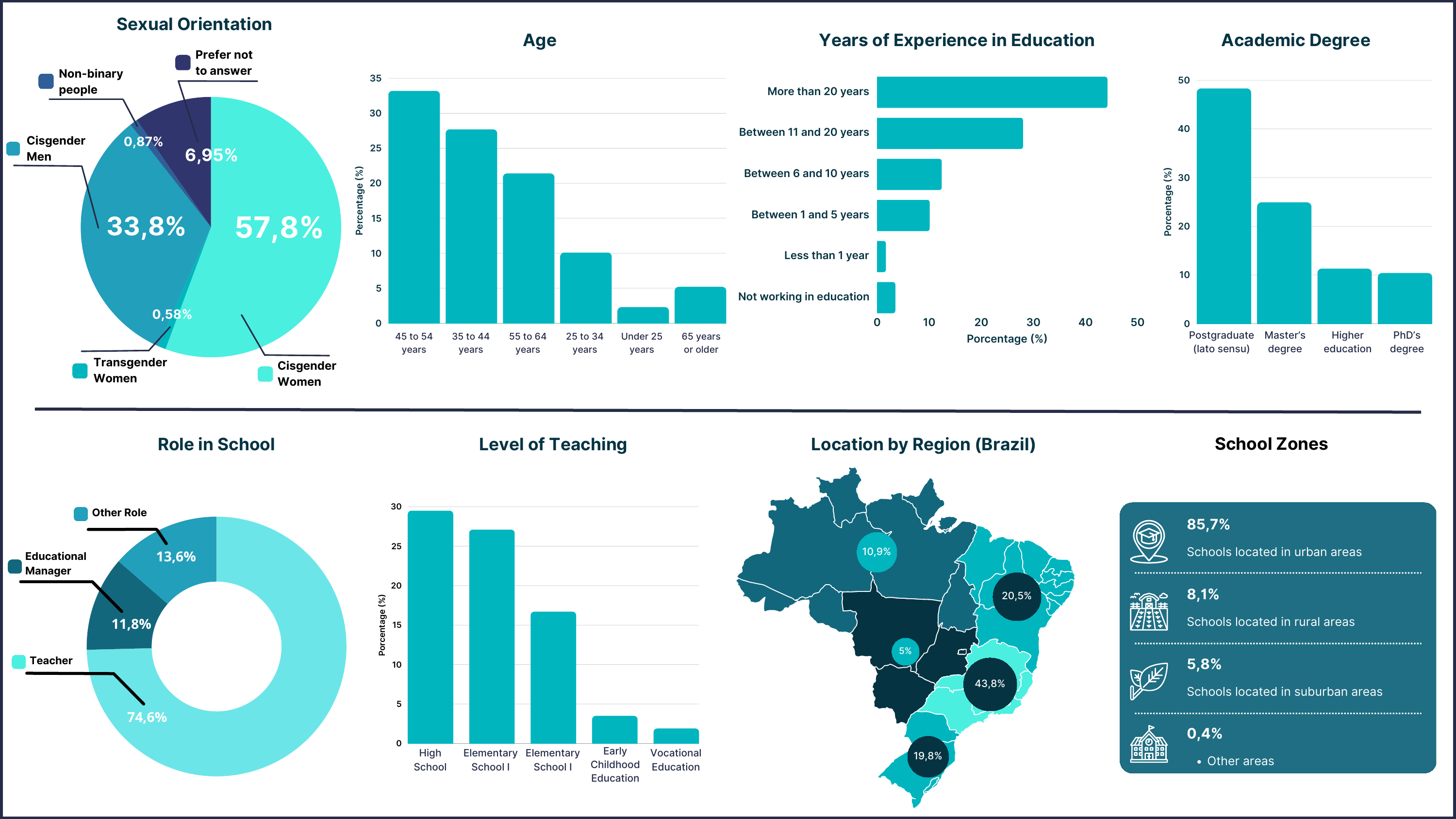}
    \caption{Characteristics of the sample (n = 346). The chart shows the distribution of participants by occupation, gender, age group, region, educational level, years of experience, school location, and institutional affiliation. These data provide an overview of the profile of the professionals who responded to the questionnaire, highlighting the diversity of educational contexts and teaching experiences.}
    \label{fig:sample}
\end{figure}

Understanding the characteristics of the sample is essential to interpret the data in context and to recognize the potential limitations of the study. Figure \ref{fig:sample} presents some main measures and characteristics of the sample collected.

Initially, it is important to distinguish the occupations of professionals who work in schools. Approximately 74.6\% of respondents are teachers, while 11.8\% hold management positions, and 13.6\% fall under the category of ``other occupation''. Thus, the results obtained reflect not only the perceptions of teachers, but also those of other education professionals working in the school environment, providing a more comprehensive view of the problem analyzed.

Regarding gender identity, 57.8\% of the participants identified as cisgender women and 33.8\% as cisgender men, while 6.95\% preferred not to disclose their identity. The sample also includes 0.58\% transgender women and 0.87\% non-binary individuals, indicating that the study considered, even in smaller proportions, different gender identities. In general, the data point to a female majority (58.38\%) among the participants, which is consistent with the 2024 School Census \citep{inep2025censo_escolar}, also highlighting the predominance of women in Brazilian basic education, both among teachers and school administrators.

Regarding the age range of the participants, the most representative group is between 45 and 54 years old, which corresponds to 33.3\% of the sample. This is followed by the age group 35 to 44 (27.7\%) and the age group 55 to 64 (21.4\%). In parallel, the 2024 School Census \citep{inep2025censo_escolar} indicates that approximately 35\% of Brazilian teachers are between 40 and 49 years old.

Furthermore, the highest concentration of responses came from the Southeast region, with 43.8\%, possibly reflecting its higher population density, which represents 41.77\% of the Brazilian population, according to the census of 2022 \cite{ibge2022censo}. The remaining regions recorded the following participation percentages: Northeast (20.5\%), South (19.8\%), North (10.9\%), and Central-West (5\%). According to the same Census \citep{ibge2022censo}, these regions correspond to 28\%, 15\%, 6.6\%, and 8\% of the national population, respectively. These data suggest that the sample represents the geographical distribution of the respondents in the country.

Taking this into account, the five states with the highest number of participants were São Paulo, Minas Gerais, Paraná, Rio de Janeiro, and Rio Grande do Sul. These states are among the six most populous in the country, except Bahia, which is the fourth according to the census of 2022 \cite{ibge2022censo}. This fact reinforces the robustness of the sample, as it aligns with the regions with the highest population density.

Next, we analyze the academic background of the participating professionals. Approximately 48.3\% have a lato sensu postgraduate degree, 24.9\% have a master's degree, 10.4\% have a doctorate, 11.3\% reported having only a bachelor's degree, and 5.1\% declared having another type of qualification. When comparing these data with the 2024 School Census \citep{inep2025censo_escolar}, it can be observed that approximately 48\% of basic education teachers hold a postgraduate degree, whether lato sensu or stricto sensu. This percentage is significantly lower than the 84.5\% of professionals with postgraduate degrees identified in this study, indicating a possible sample bias, potentially related to the profile of the respondents.

Another relevant aspect concerns the length of experience in the field of education. The majority of the participants (44.2\%) have more than 20 years of experience, followed by 28.1\% with 11 to 20 years of experience. Additionally, 12.4\% reported having 6 to 10 years of experience, while 10.1\% are in the range of 1 to 5 years. The least represented groups are those with less than 1 year of experience (1.7\%) and those who do not directly work in education (3.5\%).

The sample also provides information on the level of education at which professionals work. Approximately 29.5\% of the respondents are associated with high school, 27.1\% work in Elementary School I (grades 1–5, and 16.7\% in Elementary School II (grades 6–9). These data help to understand the diversity of educational contexts represented in the study. 

Finally, we analyzed the location of the educational institutions where professionals work. Most of the schools are located in urban areas, accounting for 85.7\% of the sample. Following this, 8.1\% of schools are located in rural areas, 5.8\% in peri-urban regions, and 0.4\% were classified as 'other'. These data indicate that, although there is a predominance of urban institutions, the study also included schools located in non-exclusively urban contexts.

We also analyze the affiliation of the institutions with the government. Public schools in the state represent the majority of the survey responses, accounting for 37.6\% of the total, followed by public schools in the municipal district 28.7\%. In third place are private institution educators, comprising 17.1\% of respondents. The participation of professionals affiliated with federal public schools is 9.3\%, while those working in universities represent 3.5\%. Furthermore, 3.8\% of the participants reported working in other contexts.

Understanding this diversity is essential, as teachers' realities vary significantly depending on the type of institution and its location. The administration, whether municipal, state, federal or private, directly influences access to infrastructure, technologies, technical support, and available pedagogical resources.

Municipal and state schools, for example, often face budgetary constraints and high demands for inclusion. In contrast, private institutions have better material conditions and greater access to technological innovations. Ignoring these disparities can result in the proposal of generic and ineffective solutions that do not consider the real and specific needs of each educational context.

%In summary, the sample is composed predominantly of teachers and shows a female majority (58.38\%). Most of the participants are over 35 years old (82.4\%), and the regional distribution of the respondents generally reflects the distribution of the population of the country. The sample also demonstrates a high level of education, with the majority of professionals holding postgraduate degrees (84.5\%), as well as extensive experience in the field, with more than 11 years of service (72.3\%). Respondents work across different stages of basic education, and are predominantly employed in institutions located in urban areas (85.7\%), although non-urban contexts are also represented.

%%%%%%%%%%%%%%%%%%%%%%%%%%%%%%%%%
%%%%%%%%%%%%%%%%%%%%%%%%%%%%%%%%%

\section{Results and Discussions}

The main findings of the study reveal a promising scenario, and the majority of educators (97.3\%) show interest in the potential of AI for teaching and learning. However, the study also points to significant challenges for its effective implementation, including the need for teacher training, the lack of adequate infrastructure, and the absence of technical support, among others.  

The initial scenario shows that most educators have basic knowledge of AI (80.3\%), indicating a limitation in using this technology beyond simple tasks and hindering their role as guiding agents on the topic. On the other hand, more than 80\% of teachers expressed interest in participating in AI continuous professional development programs, with a preference for online courses that include videos, quizzes, and collaborative spaces.

This indicates that educators are open to learning about the topic, but underscores the need for access to structured and practical materials, which eliminate the need for the teacher to search for, curate, and pre-evaluate quality content. In this context, it becomes essential to consider the particularities of the reality of each teacher and the area of instruction, since generalist courses often do not address these specificities, resulting in training that lacks practical applicability and effectiveness in daily teaching practice. Taking into account the different dimensions presented previously, and the main points observed, the scenario resulting from this analysis is presented below. Some results are shown in Figure \ref{fig:result}.
%In this section, the responses collected in this study will be discussed, organized according to the different dimensions presented previously. Some of the information is shown in Figure \ref{fig:result}.

\begin{figure}[!htp]
    \centering
    \includegraphics[width=0.95\linewidth]{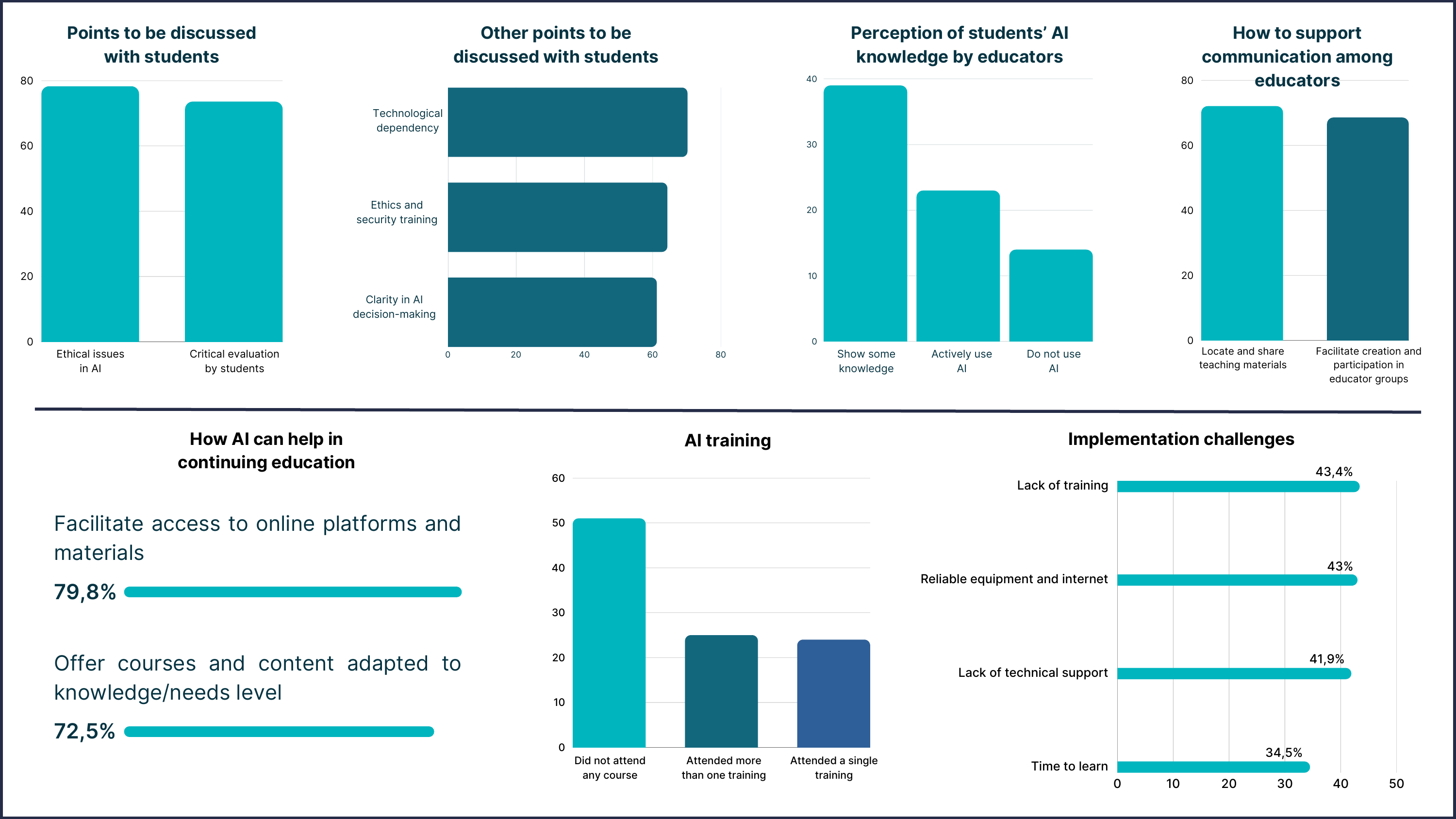}
    \caption{Summary of respondents’ perceptions regarding AI in education (n=346). The figure illustrates the participants’ knowledge, perceived benefits, and engagement with AI across three dimensions.}
    \label{fig:result}
\end{figure}

\subsection{Teaching and Learning with Digital Technologies}

We observed that most of the respondents have limited knowledge on the subject: approximately 69.4\% reported having a basic understanding, 9.7\% stated they had only heard about the topic, and 1.2\% indicated not knowing at all. In contrast, 17.8\% reported having advanced proficiency and 1.9\% consider themselves experts. 

Next, aligned with the competencies of the first dimension, we analyze the participants' perceptions about the potential benefits of AI in teaching. In this question, respondents who recognized the potential of AI could select multiple areas in which it could contribute, while those who did not identify benefits selected only that single option. Considering this, 97.3\% of the participants indicated at least one positive aspect among those listed.  

The options cited most frequently were the creation of interactive content (80.6\%) and lesson planning (80.2\%). Other benefits mentioned included personalized assessment (68.6\%), the potential for inclusion (65.5\%), and the identification of students’ learning difficulties (47.7\%). The study also revealed that teachers are interested in using AI for the analysis of educational data, perceiving it as a valuable resource for pedagogical practices. 

In this regard, 79.8\% of the participants highlighted the identification of learning trends as the main benefit. In the field of inclusive practices, 71.3\% pointed to the promotion of personalized learning strategies, while 70.9\% recognized the importance of integrating assistive technology resources. Other benefits mentioned include the creation of collaborative learning environments (60.5\%) and the monitoring of student progress, with a focus on identifying areas that require special attention (53.5\%). Only 0.8\% stated that they do not believe in the potential of AI to promote inclusion.

Furthermore, the majority of respondents consider that AI can optimize the time dedicated to activities such as researching relevant content (80.6\%) and creating assessments and exercises (70.9\%) without compromising pedagogical quality. Other positively perceived impacts include the curation of digital content (66.3\%) and the remixing or adaptation of materials (51.2\%). These results reinforce the perception of AI as a supportive tool for the production of teaching materials, capable of increasing teacher productivity and freeing up more time for higher-value pedagogical activities.

\subsection{Digital Citizenship}

Moving on to the second dimension, digital citizenship, the results indicate that 78.3\% of educators consider it essential to promote students' education on ethical issues related to AI, recognizing this aspect as crucial for the critical use of these tools. Furthermore, 73.6\% emphasized the importance of teaching students how to critically evaluate AI-generated information.

Other aspects were also widely highlighted by the participants. Among them, 70.2\% emphasized the importance of reflecting on technological dependence and its impacts in the classroom. Furthermore, 64.3\% considered it essential to ensure that teachers receive adequate training in ethics and safety in the use of AI. Finally, 61.2\% stressed the need to ensure transparency in decisions made by AI systems, as well as in the identification and mitigation of their biases.

Regarding the responsible use of AI by students, both inside and outside the classroom, 39.1\% of teachers reported that students demonstrated some knowledge of the topic but did not use these tools frequently. Meanwhile, 22.5\% indicated that students actively use AI in their daily academic and academic activities. In contrast, 13.6\% reported that most students do not use AI in their practices.

\subsection{Professional Development}

The third dimension refers to professional development. In this area, 79.8\% of teachers highlighted the importance of facilitating access to online platforms and learning materials about AI. Furthermore, 72.5\% indicated that the provision of courses and content tailored to individual levels and knowledge needs is relevant. Other important aspects include the suggestion of courses and training aligned with the professional trajectory (65.5\%), the promotion of collaboration and experience exchange between educators (65.1\%), and the possibility of using AI for performance analyses to identify areas for development (57.4\%). Only 0.4\% of respondents do not believe that AI can contribute to continuous professional development.  

However, the majority of respondents (50.8\%) stated that they have not participated in any AI courses in education but are interested. Another 25.2\% reported having participated in more than one training session, while 23.6\% attended a single course, and 0.4\% expressed no interest. Teachers also recognize that AI can facilitate communication and collaboration, as well as support professional development. The main highlight was the use of AI to locate and share teaching materials available online, indicated by 72.1\% of the respondents. Other benefits mentioned include facilitating the creation and participation in educator groups (68.6\%) and supporting the co-creation of digital resources and educational content.  

Regarding the level of support and encouragement for the use of AI, the majority (31.8\%) believe that the school offers some encouragement but is still insufficient. Another 26.4\% indicate that the school leadership and the school community are aware but do not promote its use, while 26.0\% report that there is no significant technology-oriented culture. Only 13,0\% reported strong leadership support, and 2.8\% were unable to respond.

\subsection{Inclusion through AI}

When discussing benefits, the most recurrent responses indicated the creation of interactive content (80.6\%) and lesson planning (80.2\%) as the main advantages. Personalized assessment (68.6\%), the potential for inclusion (65.5\%), and the identification of students’ learning difficulties (47.7\%) were also highlighted. Regarding inclusion, the study also indicated that the respondents recognize the potential of AI as an ally in creating teaching materials that are more accessible and compatible with the diverse needs of students, with the content personalization proposal widely acknowledged in other studies \citep{art_ger, opport_canada}.

According to the 2024 School Census \citep{inep2025censo_escolar}, special education enrollments exceeded 2.1 million, representing an increase of 58.7\% in the past five years. This growth reinforces the need highlighted by study participants for the creation of interactive content (80.6\%), personalized activities (84.1\%), and the development of accessible resources (81\%), demonstrating the importance of inclusive strategies supported by technologies such as AI.

To contextualize this scenario, we can analyze student-to-class ratios from the 2024 School Census \citep{inep2025censo_escolar} and teacher working hours, according to the OECD's Education at a Glance 2024 report \citep{oecd2024education}. National data indicate that the average number of students per class is 15.4 in early childhood education, 20.9 in the lower grades of primary education, 25.9 in the upper grades, and 28.9 in secondary education. These numbers show that class sizes tend to increase with the education level, requiring teachers to manage greater diversity, and, consequently, leaving less time to address individual student needs.

Regarding workload, the OECD report indicates that Brazilian teachers work on average 800 hours per year. Although these data are not detailed by education level for Brazil, it is observed that, compared to the average of OECD countries, Brazil only has a lower workload in the pre-primary education segment. The OECD averages for pre-primary, primary, lower secondary, and upper secondary education are 1,007, 773, 706, and 679 annual hours, respectively. It should be noted that these metrics include both classroom time and other professional activities, which requires caution in interpretation. However, the data suggest that, on an international scale, Brazilian teachers have a higher workload than the average at several education levels.

In this context, a tool that provides support to educators is likely to be highly accepted, as it would contribute not only to personalized teaching but also to alleviating teacher workload. With large class sizes and high annual working hours, many teachers face difficulties in meeting the diverse needs of students, making the support of solutions that facilitate adaptation to the heterogeneous realities of classrooms essential. Unfortunately, the inclusion of AI in schools presents several challenges, according to teachers, such as lack of training (43.4\%), reliable equipment and internet access (43.0\%), technical support (41.9\%), and time to learn to use new tools (34.5\%). These factors highlight structural and support-related problems present in our educational institutions.

%%%%%%%%%%%%%%%%%%%%%%%% RECORTE MUNDO - abordagens governamentais %%%%%%%%%%%%%%%%%%%%%%%%%%%%%%%

\subsection{Comparative Perspectives on AI Education Policies}

In the Brazilian context, only 48.8\% of teachers reported having participated in any type of course on the topic. When these data are compared with Canada, based on the study by \cite{opport_canada}, about 63\% of a sample of 76 teachers reported having received some technical training on the subject, indicating that Brazil has a smaller number of trained teachers. However, this comparison lacks more detailed information, such as the level and duration of training in each country. In parallel, the study mentioned reached results similar to those found here, where the main concern of educators is the lack of training and technical support for the adoption of this technology.

In the same context, \cite{opport_canada} defines GAI integration using a bottom-up approach, where there is a lack of clear guidelines, and students, parents, and teachers do not have an official framework for learning and implementation of AI in schools, as occurs in Canada. In contrast, in other locations, such as Hong Kong, there is an official AI curriculum that promotes responsible use in classrooms.  According to the same author, a division is beginning to emerge in AI policy: some regions adopt a top-down approach, developing safeguards and official frameworks for responsible teaching, while other regions are still in a bottom-up process with decentralized adoption, gathering evidence to support AI use. In this scenario, a decentralized approach results in a lack of standardization in education, where some programs may be well-structured with trained teachers, while others are less organized, which does not ensure a minimum understanding of safety, ethics, and bias related to AI across society.

In these terms, Brazil is still positioned within a bottom-up model, as even though the \textit{"Referencial de Saberes Digitais Docentes"} \citep{brasil2023referencial} encompasses some AI-related topics, and the ``Brazilian Artificial Intelligence Plan (PBIA) 2024-2028'' \citep{plano_ia} defines investment guidelines for AI, including in education, there is still no official curriculum specifying what should be covered, or official courses to ensure adequate training of students and teachers across the entire country. Nevertheless, some states, such as Piauí, have introduced an AI course, being the first territory in the Americas to implement such a subject in basic education \citep{governopi2024unesco}. 

%%%%%%%%%%%%%%%%%%%%%%%% ETICA E RESPONSABILIDADE SOCIAL   %%%%%%%%%%%%%%%%%%%%%%%%%%%%%

\subsection{Ethics and Social Responsibility}

Consequently, the lack of official guidelines affects ethics training and responsible use of this technology, a central concern of 78.3\% of the respondents in this study. In this context, being aware of the existence of these tools, their biases, and ethical implications enables students to make informed decisions. A relevant example concerns DeepFakes, which can be understood as the synthetic generation of audiovisual media \cite{deepfake}. Despite its benefits, this technology is predominantly associated with negative uses, such as political manipulation to influence the democratic process \citep{broinowski2024beyond}, and is generally recognized as capable of facilitating disinformation and preventing access to authentic sources \citep{art_ger}. Unlike other media creation technologies, DeepFakes stands out because it requires a fraction of the cost and computational time, and it can be produced by anyone with access to the appropriate AI tools \citep{broinowski2024beyond}. However, detecting fake DeepFake videos is not straightforward, depending on the quality of the material, with detection still being unreliable and difficult to scale \citep{broinowski2024beyond}.

In parallel, in Brazil, according to the 2024 Household Sample Survey Coordination \citep{ibge2025internet}, 89.1\% of Brazilians over 10 years old have internet access, primarily (95\%) for video and voice calls, with 84.2\% reporting use of social networks, and only 68.9\% using the internet to "read newspapers, news, books, or magazines". Based on these data, 75\% of Brazilians over 10 years of age access social networks, potentially exposed to AI-generated content but not necessarily able to discern whether it is real or not, since they may not even know that some content can be artificially generated, which may influence their perception of reality for political or other purposes. All of this highlights the need to bring ethical discussions that occur in society into schools, creating an environment that allows students to understand their reality. Clearly, educating students does not solve the problem of DeepFakes, but equips them with knowledge of the existence of these tools and enables them to make use of the beneficial aspects of such technologies.

%%%%%%%%%%%%%%%%%%%%%%%%%   PROBLEMAS NA INFRAESTRUTURA      %%%%%%%%%%%%%%%%%%%%%%%

\subsection{Challenges in Brazilian Infrastructure}

It is essential to discuss the benefits that AI can bring to education, but it is also necessary to understand the practical obstacles that hinder its implementation. The lack of training and time to learn how to use these tools is directly related to the overloaded environment in which educators operate, as previously mentioned. Furthermore, the absence of technical support, reported by 41.9\% of the participants, represents a significant barrier, referring to the shortage of professionals in schools who can assist with technical issues involving computers, projectors, digital whiteboards, and even internet connectivity.

Nevertheless, the challenges go beyond training, encompassing structural problems in schools that limit the implementation of technologies. The 2024 School Census \citep{inep2025censo_escolar} indicates that 85.9\% of basic education schools have broadband internet access. However, the Connected Education program \citep{brasil_educacaoconectada}, which evaluates the quality of connection, reveals regional disparities. In the Southeast region, considered the one with the best indicators, only a portion of schools have Internet with adequate speed. For example, in São Paulo, 4,964 of 19,006 schools have installed satisfactory measurement devices and connections, corresponding to a connectivity rate of 26.1\%. Minas Gerais leads the region with 30.3\%, followed by Espírito Santo (29.4\%) and Rio de Janeiro (24.4\%). These data show that, although most schools have internet access, the quality, and adequacy of this access are still insufficient to ensure full use of digital technologies.

By analyzing other aspects of infrastructure in primary and secondary education based on the 2024 School Census \citep{inep2025censo_escolar}, we gain a better understanding of the scenario and the limitations of using technology in the classroom. Only 70\% of schools have multimedia projectors, while around 21\% have digital whiteboards, highlighting a limitation in the use of resources for real-time projection of content, including AI-generated materials or the tools themselves, during lessons. Regarding equipment, approximately 56\% of primary and secondary schools have access to desktop computers, even with a lower availability of laptops and tablets. This limitation hinders the execution of practical activities using digital tools, where students could explore, under teacher supervision, aspects related to AI.

%%%%%%%%%%%%%%%%%%%%%%%%%%   CONSIDERAÇÕES FINAIS  %%%%%%%%%%%%%%%%%%%%%%%%

\subsection{Literature Recommendations and the Brazilian Context}

The implementation of AI and GAI in Brazilian basic education requires careful alignment between state-of-the-art discussions, the specificities of the national context, and educators' perceptions. Considering these three pillars is essential to ensure that technologies are used effectively, inclusively, and ethically, while respecting local structural limitations and potentials. Based on this study and evidence from the literature, we present recommendations to guide policies and practices that can enable the productive use of AI in the Brazilian school environment.

\subsubsection{Continuing Education and Contextualized Teacher Training}

It is essential to develop online, free, and easily accessible training programs that enable teachers to integrate AI into the creation of personalized teaching materials, lesson planning, and adaptive assessment. The initial choice of an online format allows a single standardized course to be offered nationwide, ensuring that all teachers receive a solid and consistent foundation, with a uniform minimum understanding of the technologies and their applications.

Furthermore, the program should take into account the varying levels of technological familiarity among educators and include specific modules on ethics, digital safety, biases, and limitations of AI models. Although other forms of training, such as in-person workshops, discussion groups, and collaborative sessions, are important and welcome, they require greater local engagement and a more complex implementation structure. Therefore, starting with an online course delivered by specialists can ensure immediate and effective progress in teacher training, serving as a foundation to later explore complementary approaches that promote the exchange of experiences and mutual support among educators.

\subsubsection{Incorporating Ethics and Digital Citizenship into the Curriculum}

AI should be used not only as a pedagogical support tool but also as a subject of study, fostering the development of students' critical thinking regarding the responsible use of technology. The curriculum must incorporate discussions on the social, ethical, legal, and emotional impacts of AI, focusing on the importance of transparency, bias mitigation, and privacy protection. Consequently, teacher training should prepare educators to facilitate these reflections, enabling them to develop conscious and critical students for responsible participation in the digital society.

%%%%%%%%%%%%%%%%%%%%%%%%%%%%%
%%%%%%%%%%%%%%%%%%%%%%%%%%%%%

\subsubsection{Investment in Infrastructure and Technical Support}

Despite enthusiasm for the potential of AI, effective implementation faces significant structural challenges, such as insufficient equipment, lack of stable internet connectivity, and lack of qualified technical support. Public policies must allocate resources to improve these conditions, ensuring that technology can be used consistently and continuously, especially in public schools in less advantaged regions. Initially, the teacher can incorporate the use of AI in the creation of teaching materials, which make up the lesson plans written on the board. At the same time, slide projection devices can be used for students; this approach does not require Internet access and is supported by the fact that approximately 75\% of schools have multimedia projectors. In this way, AI-generated audiovisual content can also be included in the teacher's instructional materials.

Alternatively, the use of computers with AI programs for teacher-supervised lessons is a possibility that expands the pedagogical potential. However, access to these resources is still limited in Brazilian schools, as individual activities require one computer per student, in addition to internet connectivity, depending on the nature of the activity. As mentioned above, only 56\% of schools have adequate computers, 30\% have adequate internet, and the overlap of these two resources may be present in an even smaller fraction of institutions. 

An alternative to introduce students to AI concepts is the use of unplugged activities, which do not require hardware or software and are effective in conveying essential computing concepts \cite{atividades_desp, atividades_desp2}. This type of approach employs physical and tangible materials to support hands-on learning, allowing exploration of fundamental ideas in the field, such as data representation, feature extraction, and rule-based decision-making \cite{atividades_desp}.  

These recommendations synthesize an integrated and pragmatic vision for the introduction of AI in Brazilian basic education, balancing educators' expectations, technological advances, and the real conditions of the educational system. Responsible and contextualized adoption has the potential to transform teaching practices, expand learning opportunities, and prepare students for the challenges of the 21st century. 

Therefore, the incorporation of AI into education presents significant potential to personalize teaching and support teachers, especially given the increasingly large class sizes and teacher workload. However, for these benefits to be effectively realized, it is essential to overcome structural challenges, such as the limited technological infrastructure of schools and the socioeconomic inequality that affects a large part of the Brazilian population. Teacher training and equitable access to digital tools constitute important pillars for the effective implementation of AI in the school environment. Finally, our research complements studies such as the School Census and other surveys, highlighting teachers' openness and recognition of the importance of including AI in the educational system, particularly as a tool to support teaching practices.

%%%%%%%%%%%%%%%%%%%%%%%%%%%%%
%%%%%%%%%%%%%%%%%%%%%%%%%%%%%

\section{Conclusions}

%Aqui podemos ser brutos hehe.

In our view, the expression 'AI for all' means that every individual, regardless of geographical location, financial resources, or prior knowledge, should have access to AI and its benefits. Motivated by the rapid emergence of guidelines and discussions on the topic, we sought to understand how this principle translates to the reality of Brazilian K–12 education and, importantly, to listen to those who will directly face the challenges and opportunities of integrating AI into classrooms: teachers themselves. To achieve this, we conducted a nationwide survey with 346 educators from all Brazilian regions.

Our findings reveal that although most teachers have only basic knowledge of AI, there is a strong interest in adopting it as a pedagogical resource, particularly to reduce workload, provide personalized support to students, and enhance teaching efficiency. Nonetheless, enthusiasm alone is not enough. The meaningful integration of AI into Brazilian education requires the elimination of substantial structural challenges, including technological limitations in schools, unequal access to digital resources, and socioeconomic disparities that continue to shape the learning environment.

Therefore, the successful and equitable adoption of AI in education depends on coordinated actions: continuous and accessible teacher training, investment in technological infrastructure, and public policies aligned with local needs. Taken together, our study highlights the current perceptions of teachers and reinforces the urgency of designing strategies that enable an inclusive, ethical, and effective digital transformation, ensuring that the promise of 'AI for all' truly becomes a reality for Brazilian students and educators.

%%%%%%%%%%%%%%%%%%%%%%%%%%%%%
%%%%%%%%%%%%%%%%%%%%%%%%%%%%%

\section{Partnership and Data Availability}

The survey conducted among teachers was carried out in collaboration between the Federal University of Technology – Paraná (UTFPR), the InteliGente Project, and the Instituto Significare. The complete report can be accessed at https://encurtador.com.br/uRrI.

%%%%%%%%%%%%%%%%%%%%%%%%%%%%%
%%%%%%%%%%%%%%%%%%%%%%%%%%%%%

\section{Acknowledgements}

This work has been funded by The Spencer Foundation under the Grant Agreement 202600243; Canadian International Development Research Centre (IDRC) under the Grant Agreement 109981, funded by IDRC and the UK government’s Foreign, Commonwealth and Development Office. Bruno Florentino has been funded by the São Paulo Research Foundation (FAPESP), grant \#2024/00830-8. Dr. Bonidia also received funding from the Regional Fund for Digital Innovation in Latin America and the Caribbean (FRIDA-LACNIC).

%%%%%%%%%%%%%%%%%%%%%%%%%%%%%
%%%%%%%%%%%%%%%%%%%%%%%%%%%%%

\bibliographystyle{unsrtnat}
\bibliography{references} 

@article{atividades_desp2,
  title={A Structured Unplugged Approach for Foundational AI Literacy in Primary Education},
  author={Carrisi, Maria Cristina and Marras, Mirko and Vergallo, Sara},
  journal={arXiv preprint arXiv:2505.21398},
  year={2025}
}

@article{atividades_desp,
title={Unplugged K-12 AI Learning: Exploring Representation and Reasoning with a Facial Recognition Game},
volume={38}, url={https://ojs.aaai.org/index.php/AAAI/article/view/30376}, 
DOI={10.1609/aaai.v38i21.30376}, 
number={21}, 
journal={Proceedings of the AAAI Conference on Artificial Intelligence}, 
author={Lim, Hansol and Min, Wookhee and Vandenberg, Jessica and Cateté, Veronica and Mott, Bradford}, 
year={2024}, 
month={Mar.}, 
pages={23285-23293} 
}

@INPROCEEDINGS{deepfake,
  author={Mary, Amala and Edison, Anitha},
  booktitle={2023 International Conference on Control, Communication and Computing (ICCC)}, 
  title={Deep fake Detection using deep learning techniques: A Literature Review}, 
  year={2023},
  volume={},
  number={},
  pages={1-6},
  doi={10.1109/ICCC57789.2023.10164881}}

@INPROCEEDINGS{norte_sul2,
  author={Chakraborty, Sourojeet and Galatro, Daniela},
  booktitle={2025 IEEE Global Engineering Education Conference (EDUCON)}, 
  title={Artificial Intelligence (AI) Equality in Engineering Education: Strategies to Unite the AI Gap between the Global North \& South}, 
  year={2025},
  volume={},
  number={},
  pages={1-10},
  doi={10.1109/EDUCON62633.2025.11016563}}

@article{opport_canada,
title = {Opportunities, challenges and school strategies for integrating generative AI in education},
journal = {Computers and Education: Artificial Intelligence},
volume = {8},
pages = {100373},
year = {2025},
issn = {2666-920X},
doi = {https://doi.org/10.1016/j.caeai.2025.100373},
url = {https://www.sciencedirect.com/science/article/pii/S2666920X2500013X},
author = {Davy Tsz Kit Ng and Eagle Kai Chi Chan and Chung Kwan Lo},
}

@ARTICLE{art_ger,
  author={Mittal, Uday and Sai, Siva and Chamola, Vinay and Sangwan, Devika},
  journal={IEEE Access}, 
  title={A Comprehensive Review on Generative AI for Education}, 
  year={2024},
  volume={12},
  number={},
  pages={142733-142759},
  doi={10.1109/ACCESS.2024.3468368}}

@misc{brasil2023referencial,
  author       = {{Brasil}},
  title        = {Referencial de Saberes Digitais Docentes},
  year         = {2023},
  howpublished = {\url{https://www.gov.br/mec/pt-br/escolas-conectadas/20240822MatrizSaberesDigitais.pdf}},
  note         = {Ministério da Educação, Secretaria de Educação Básica. Brasília. Accessed: 6 Aug. 2025}
}

@misc{plano_ia,
  author       = {{Brasil}},
  title        = {Plano Brasileiro de Inteligência Artificial 2024--2028},
  year         = {2024},
  howpublished = {\url{https://www.gov.br/lncc/pt-br/assuntos/noticias/ultimas-noticias-1/plano-brasileiro-de-inteligencia-artificial-pbia-2024-2028}},
  note         = {Ministério da Ciência, Tecnologia e Inovação. Brasília. Accessed: 6 Aug. 2025}
}

@misc{governopi2024unesco,
  author       = {{Governo do Piauí}},
  title        = {UNESCO reconhece Piauí como primeiro território nas Américas a implementar o ensino de Inteligência Artificial na educação básica},
  year         = {2024},
  howpublished = {\url{https://www.pi.gov.br/unesco-reconhece-piaui-como-primeiro-territorio-nas-americas-a-implementar-o-ensino-de-inteligencia-artificial-na-educacao-basica/}},
  note         = {Accessed: 5 Aug. 2025}
}

@misc{inep2025censo_escolar,
  author       = {{BRASIL. INEP}},
  title        = {Censo Escolar da Educação Básica 2024: Resumo Técnico},
  year         = {2025},
  address      = {Brasília},
  howpublished = {Instituto Nacional de Estudos e Pesquisas Educacionais Anísio Teixeira (Inep)},
  url =  {https://www.gov.br/inep/pt-br/areas-de-atuacao/pesquisas-estatisticas-e-indicadores/censo-escolar/resultados}
}

@article{HKTDC2025Brazil,
  title        = {Latin America: Brazil’s Economic Development},
  author       = {Tsang, Alice and Fu, Nicholas},
  year         = {2025},
  month        = {April},
  journal      = {HKTDC Research},
  url          = {https://research.hktdc.com/en/article/MTk5MTYxNDg1Mg},
  note         = {Accessed: October 22, 2025}
}

@book{oecd2024education,
  author    = {{OECD}},
  title     = {Education at a Glance 2024: OECD Indicators},
  year      = {2024},
  publisher = {OECD Publishing},
  address   = {Paris},
  doi       = {10.1787/c00cad36-en},
  url       = {https://doi.org/10.1787/c00cad36-en},
}

@misc{brasil_educacaoconectada,
  author       = {{BRASIL. Ministério da Educação}},
  title        = {Programa de Inovação Educação Conectada},
  year         = {2025},
  address      = {Brasília, DF},
  howpublished = {\url{https://medidor.educacaoconectada.mec.gov.br}},
  note         = {Acess in: 1 abr. 2025}
}

@article{broinowski2024beyond,
  author  = {Broinowski, A. and Martin, F. R.},
  title   = {Beyond the deepfake problem: Benefits, risks and regulation of generative AI screen technologies},
  journal = {Media International Australia},
  volume  = {0},
  number  = {0},
  year    = {2024},
  doi     = {10.1177/1329878X241288034},
  url     = {https://doi.org/10.1177/1329878X241288034}
}

@book{ibge2025internet,
  author       = {{BRASIL. Instituto Brasileiro de Geografia e Estatística (IBGE)}},
  title        = {Acesso à internet e à televisão e posse de telefone móvel celular para uso pessoal 2024},
  year         = {2025},
  address      = {Rio de Janeiro},
  publisher    = {IBGE},
  note         = {Coordenação de Pesquisas por Amostra de Domicílios. Disponível somente em meio digital. Acess in: 1 abr. 2025},
  isbn         = {9788524046636},
  series       = {Coleção Ibgeana},
}

@book{ibge2022censo,
  author       = {{BRASIL. Instituto Brasileiro de Geografia e Estatística (IBGE)}},
  title        = {Censo Demográfico 2022},
  year         = {2022},
  publisher    = {IBGE},
  address      = {Rio de Janeiro},
  note         = {Disponível em: \url{https://cidades.ibge.gov.br/brasil/pesquisa/10102/122229}. Acess in: 5 ago. 2025}
}

@ARTICLE{transforming_ai_benefits,
  author={Lang, Qi and Wang, Minjuan and Yin, Minghao and Liang, Shuang and Song, Wenzhuo},
  journal={IEEE Transactions on Learning Technologies}, 
  title={Transforming Education With Generative AI (GAI): Key Insights and Future Prospects}, 
  year={2025},
  volume={18},
  number={},
  pages={230-242},
  doi={10.1109/TLT.2025.3537618}}

@INPROCEEDINGS{zhang_ai_benefits,
  author={Zhang, Jing and Sun, Di},
  booktitle={2025 7th International Conference on Computer Science and Technologies in Education (CSTE)}, 
  title={A Systematic Review of Generative Artificial Intelligence in Education}, 
  year={2025},
  pages={552-556},
  doi={10.1109/CSTE64638.2025.11092288}}

@inproceedings{qiu_ai_benefits,
author = {Qiu, Xiaoming and Zhang, Shu},
title = {Application analysis of generative artificial intelligence in basic education},
booktitle={Proceedings of the 2024 7th International Conference on E-Business, Information Management and Computer Science},
year = {2025},
isbn = {9798400712876},
publisher = {Association for Computing Machinery},
address = {New York, NY, USA},
url = {https://doi.org/10.1145/3723420.3723428},
doi = {10.1145/3723420.3723428},
pages = {41–46},
numpages = {6},
series = {EBIMCS '24}
}

@Article{berg_ai_benefits,
AUTHOR = {van den Berg, Geesje and du Plessis, Elize},
TITLE = {ChatGPT and Generative AI: Possibilities for Its Contribution to Lesson Planning, Critical Thinking and Openness in Teacher Education},
JOURNAL = {Education Sciences},
VOLUME = {13},
YEAR = {2023},
NUMBER = {10},
ARTICLE-NUMBER = {998},
URL = {https://www.mdpi.com/2227-7102/13/10/998},
ISSN = {2227-7102},
DOI = {10.3390/educsci13100998}
}

@article{gozzi2024bridging,
  author  = {Gozzi, N. and Comini, N. and Perra, N.},
  title   = {Bridging the digital divide: mapping Internet connectivity evolution, inequalities, and resilience in six Brazilian cities},
  journal = {EPJ Data Science},
  volume  = {13},
  number  = {69},
  year    = {2024},
  doi     = {10.1140/epjds/s13688-024-00508-8},
  url     = {https://doi.org/10.1140/epjds/s13688-024-00508-8}
}

@misc{portulans2024nri,
  author       = {{Portulans Institute}},
  title        = {Network Readiness Index 2024: Brazil Country Profile},
  year         = {2024},
  howpublished = {\url{https://download.networkreadinessindex.org/reports/countries/2024/brazil.pdf}},
  note         = {Acess in: 6 aug. 2025}
}

@misc{global_north,
  author       = {Yu, D. and Rosenfeld, H. and Gupta, A.},
  title        = {The {AI} Divide between the Global North and the Global South},
  year         = {2023},
  howpublished = {\url{https://www.weforum.org/agenda/2023/01/davos23-ai-divide-global-north-global-south/}},
  note         = {Accessed: 6 Aug. 2025},
  organization = {World Economic Forum}
}
%%%%%%%%%%%%%%%%%%%%%%%%%%%%%
%%%%%%%%%%%%%%%%%%%%%%%%%%%%%
\end{document}